\documentclass{aa}  
\usepackage{graphicx}
\usepackage{txfonts}
\usepackage{natbib}
\bibpunct{(}{)}{,}{a}{}{,}
\begin{document}
\newcommand{\doceCO}{\mbox{$^{12}$CO}}
\newcommand{\doce}{\mbox{$^{12}$CO}}
\newcommand{\trece}{\mbox{$^{13}$CO}}
\newcommand{\treceCO}{\mbox{$^{13}$CO}}
\newcommand{\jsc}{\mbox{$J$=6$-$5}}
\newcommand{\jtd}{\mbox{$J$=3$-$2}}
\newcommand{\jcc}{\mbox{$J$=5$-$4}}
\newcommand{\jdu}{\mbox{$J$=2$-$1}}
\newcommand{\juc}{\mbox{$J$=1$-$0}}
\newcommand{\gsim}{\raisebox{-.4ex}{$\stackrel{>}{\scriptstyle \sim}$}}
\newcommand{\lsim}{\raisebox{-.4ex}{$\stackrel{<}{\scriptstyle \sim}$}}
\newcommand{\psim}{\raisebox{-.4ex}{$\stackrel{\propto}{\scriptstyle \sim}$}}
\newcommand{\kms}{\mbox{km~s$^{-1}$}}
\newcommand{\s}{\mbox{$''$}}
\newcommand{\mloss}{\mbox{$\dot{M}$}}
\newcommand{\my}{\mbox{$M_{\odot}$~yr$^{-1}$}}
\newcommand{\ls}{\mbox{$L_{\odot}$}}
\newcommand{\ms}{\mbox{$M_{\odot}$}}
\newcommand{\mm}{\mbox{$\mu$m}}
\def\arcdeg{\hbox{$^\circ$}}
\newcommand{\seca}{\mbox{\rlap{.}$''$}}
\newcommand{\dega}{\mbox{\rlap{.}$^\circ$}}
\newcommand{\aprop}{\raisebox{-.4ex}{$\stackrel{\propto}{\scriptstyle\sf \sim}$}}
\newcommand{\apropg}{\raisebox{-.4ex}{$\stackrel{\Large \propto}{\sim}$}}
\title{CO observations of symbiotic stellar systems~\thanks{Based on
    observations carried out with the IRAM Pico Veleta 30m telescope. 
IRAM is supported by INSU/CNRS (France), MPG (Germany)
and IGN (Spain).}}

   \subtitle{}

   \author{V. Bujarrabal
          \inst{1}
	  \and
	  J. Miko\l ajewska
          \inst{2}
          \and
J. Alcolea
          \inst{3}
          \and
G. Quintana-Lacaci
          \inst{4}
         }


   \institute{   Observatorio Astron\'omico Nacional. Ap 112, E-28803 
Alcal\'a de Henares, Spain\\
              \email{v.bujarrabal@oan.es}
	      \and
N.\ Copernicus Astronomical Center, Bartycka 18, PL-00716 Warsaw, Poland \\
              \email{mikolaj@camk.edu.pl}
	      \and
Observatorio Astron\'omico Nacional (IGN), Alfonso XII N$^{\circ}$3,
              E-28014 Madrid, Spain  \\
              \email{j.alcolea@oan.es}
	      \and
Instituto de Radioastronom{\'i}a Milim\'etrica (IRAM),
Avda.\ Divina Pastora 7, E-18012 Granada, Spain \\
              \email{quintana@iram.es}
             	      }

   \date{accepted 23/02/2010}

  \abstract
   {}
   {We have studied the molecular content of the circumstellar environs
  of symbiotic stellar systems,
  in particular of the well know objects R Aqr and CH Cyg. The study of 
  molecules in these stars will help to understand the properties of the
  very inner shells around the cool stellar component, from which molecular
  emission is expected to come.
}
   {We have performed mm-wave observations with the IRAM 30m telescope
   of the \doce\ \juc\ and \jdu, \trece\ \juc\ and \jdu, and SiO \jcc\
   transitions in the symbiotic stars R Aqr, CH Cyg, and HM Sge. 
The data were analyzed
   by means of a simple analytical description of the general
   properties of molecular emission from the inner shells around the
  cool star. Numerical calculations of the expected line 
   profiles, taking into account the level population and radiative
  transfer under such conditions, were also performed.
  }
   {Weak emission of \doce\ \juc\ and \jdu\ was detected in R Aqr and
   CH Cyg; a good line profile of \doce\ \jdu\ in R Aqr was obtained. The
   intensities and profile shapes of the detected lines are compatible
   with emission coming from a very small shell around the Mira-type
   star, with a radius comparable to or slightly smaller than the
   distance to the hot dwarf companion, 10$^{14}$ --
  2\,$\times$\,10$^{14}$\,cm. 
   We argue that other possible explanations are improbable. 
   This region probably shows properties similar to those
   characteristic of the inner shells around standard AGB stars:
   outwards expansion at about 5 -- 25 \kms, with a significant
   acceleration of the gas, temperatures decreasing with
   radius between about 1000 and 500 K, and densities $\sim$ 10$^9$ --
   3\,$\times$\,10$^8$ 
   cm$^{-3}$. Our model calculations are able to explain the asymmetric
   line shape observed in \doce\ \jdu\ from R Aqr, in which the
   relatively weaker red part of the profile would result from
   selfabsorption by the outer layers (in the presence of a velocity
   increase and a temperature decrease with radius). The mass-loss rates
   are somewhat larger than in standard AGB stars, as often happens for
   symbiotic systems. In R Aqr, we find that the total mass of the CO
   emitting region is $\sim$ 2 -- 3 $\times$ 10$^{-5}$ \ms,
   corresponding to \mloss\ $\sim$ 5 $\times$ 10$^{-6}$ --  
   10$^{-5}$ \my, and compatible with results obtained from
   dust emission. Taking into account other 
   existing data on molecular emission, we suggest that the
   small extent of the molecule-rich gas in symbiotic systems is mainly
   due to molecule photodissociation by the radiation of the hot dwarf
   star.
}
{}


   \maketitle
%

\section{Introduction}

Symbiotic stellar systems (SSs) are very close binary systems composed
by a cold red giant and a very hot dwarf companion. The strong
interaction between both stars yields a number of interesting and
sometimes spectacular phenomena, like the ejection of fast collimated
flows and the formation of high-excitation, bipolar nebulae
\citep[see e.g.][and references therein]{corradi03}.  They are thus a
very attractive laboratory for studying various aspects of stellar
evolution in binary systems and of the circumstellar chemistry and
structure under these (extreme) conditions.

Some symbiotic systems show
strong IR excesses, due to the emission of dust grains formed in the
circumstellar envelopes (CSEs) ejected by the cold primary. In these
objects, the ejection of circumstellar gas from the primary is thought
to be particularly important. However, molecules are abundant only in
the innermost regions of the CSEs around SSs. Bands of CO, TiO, and
other molecules are often detected in SSs, coming from the outer
atmosphere of the cool primary, as in standard red giants. However, to
our knowledge, molecular low- and intermediate-excitation emission (SiO
masers, CO thermal lines, etc), known to come from the circumstellar
shells, had been only detected in two SSs: R Aqr and H1-36
\citep{ivison94,ivison98,schwarz95,seaquist95}. SiO maser emission
(v$>$0 lines at mm wavelengths) in R Aqr is relatively 'normal',
compared to that observed in other AGB stars
\citep[e.g.][]{pardo04,cotton04,kamohara10}. We recall that SiO maser
emission comes from the innermost circumstellar regions, in our case at
about 5 $\times$ 10$^{13}$ cm.
H$_2$O masers have been also detected in these same two objects, with
characteristics that suggest that H$_2$O emission forms only at a few
stellar radii \citep{seaquist95,ivison94,ivison98}. In R Aqr, an
H$_2$O-rich shell extending also $\sim$ 5 $\times$ 10$^{13}$ cm has been
detected from observations of H$_2$O vibrational bands
\citep{ragland08}. Molecules characteristic of the outer shells are yet
weaker and very rarely detected. OH masers have been well detected only
in H1-36 \citep{ivison94,seaquist95}. The CO thermal lines, which in
general come from still outer shells (often within about 10$^{17}$ cm),
are extremely weak in SSs. Previous to our work, only a very tentative
detection of R Aqr had been reported by \cite{groenewegen99}.
In fact, our observations (see next
sections) show that the feature detected in R Aqr by these authors is
mostly due to baseline ripples in the spectrum, since the actual
intensity of the line is about three to four times lower.

It seems well established that the detectability of molecular emission
in SSs increases for lines coming from very compact regions around the
AGB star and for systems showing relatively large distances between the
stars \citep{schwarz95,ivison98}. Therefore, the lack of detections of
molecular lines is probably due to photodissociation by the UV
radiation from the hot companion or to dynamical disruption of the
emitting regions (see further discussion in Sect.\ 5).

The properties of the innermost shells around red giants, within about
10 stellar radii, from which molecular lines in SSs seem to come, are
not very well known even for isolated stars.  Both theoretical and
observational studies
\citep[e.g.][]{hinkle82,hofner98,andersen03,sandin08} suggest that
relatively high temperatures, between 500 and 1000 K, and densities,
10$^8$ -- 10$^9$ cm$^{-3}$, are present.  These inner shells are
thought not to show the fast expansion characteristic of the outer
regions. They are probably pulsating, due to shocks originated from the
photospheric pulses, or show incipient expansion, since in these
regions dust grains are being formed and radiation pressure efficiently
acts onto them. In symbiotic systems, pulsation and outwards
acceleration may also dominate the dynamics of the inner circumstellar
layers, but the gravitational effects of the secondary cannot be
neglected, since SSs are interacting systems. In particular, it is
remarkable that SSs present mass-loss rates systematically larger than
those of isolated AGB stars \citep[e.g.][and references
therein]{miko99}, which may be due to such gravitational effects.

In this paper, we present observations of molecular mm-wave lines in
three SSs: R Aqr, CH Cyg, and HM Sge. \doce\ emission is well detected
in R Aqr and CH Cyg.

R Aqr and CH Cyg are bright and nearby symbiotic stars, extensively
studied over the whole spectral range. The distance to R Aqr has been
accurately determined from recent VLBI measurements of its parallax by
Kamohara et al.\ (2010); we will adopt their distance value, $D$ =
214$^{+45}_{-32}$. A distance of $244^{+49}_{-35}$ pc was measured for
CH Cyg from Hipparcos data \citep{vanleu07}. In both systems, the hot
component is an accreting white dwarf showing spectacular activity:
irregular accretion-powered outbursts accompanied by massive outflows
and jets (e.g.\ Kellogg et al.\ 2007, Karovska et al.\ 2007, and
references therein). The cool component in R Aqr is a Mira variable
with a pulsation period of $387^{\rm d}$,
{whereas in CH Cyg it is an M7\,III semiregular variable with a
complex variability}
\citep[e.g.][and references therein]{gromadzki09,mikolajewski92}. Both
have relatively well known orbital parameters
\citep{gromadzki09,hinkle09}.
 Aqr and CH Cyg have the longest orbital periods measured in well
studied symbiotic systems, respectively 43.6 yr and 15.6 yr.  The
orbital solution for R Aqr \citep{gromadzki09} implies that the average
component separation is $\sim$ 2.25 $\times$ 10$^{14}$ cm (15
AU). During our observations the component separation was $\sim$ 17.6
and 16.5 AU, in May 2008 and May 2009, respectively.
In the case of CH Cyg, the orbital elements from Hinkle et al.\ (2009)
yield an average component separation of $\sim$ 9 AU (1.35 $\times$
10$^{14}$ cm), and during the May 2009 observation the separation was
$\sim 9.8$ AU.  The red giant radius in both systems is known from
interferometric measurements, $\sim 1.9$ AU in R Aqr (Gromadzki \&
Miko{\l}ajewska 2009, and references therein), and $\sim 1.2$ AU in CH
Cyg (e.g. Dyck et al.\ 1998), with uncertainty set mostly by the
uncertainty in their distances, $\sim 15\, \%$ in both cases.

The symbiotic nova HM Sge is composed of a Mira variable with a
pulsation period of 527$^{\rm d}$, embedded in an optically thick dust
shell, and a white dwarf companion slowly declining from a
thermonuclear nova outburst started in 1975 (Belczynski et al.\ 2000,
and references therein; Muerset \& Nussbaumer 1994).  The orbital
period is unknown, likely higher than $\sim 100$ yr. Eyres et al.\
(2001) measured the binary component positions using HST images, and
estimated a projected binary separation of $40 \pm 9$ mas, and a
position angle of the binary axis of $130 \pm 10$ degrees, in agreement
with that suggested by Schmid et al.\ (2000) based on
spectropolarimetry.  Unfortunately, the distance to HM Sge is rather
uncertain -- published values range from 0.3 to 3.2 kpc (e.g.\ Richards
et al.\ 1999). Eyres et al.\ adopted $D$ = 1.25 kpc, and a component
separation of 50 AU. {The recently revised period-luminosity relation
for single Miras (Whitelock et al.\ 2008) would place HM Sge at $D$ =
2.5 kpc; this estimate is however uncertain due to the peculiar nature
of the object, particularly because of the low amplitude and poor
periodicity of its light curve (see the AAVSO database). }

As we will see, the CO lines in these systems are extremely weak,
typically about 100 times weaker than for standard AGB stars. We will
argue that this low intensity, as well as the other main properties of
the detected lines, show that the CO emission only comes from very
inner circumstellar regions around the cool stellar component, closer
than about 1--2 10$^{14}$ cm.

\section{Observations and data reduction}

We have used the IRAM 30m telescope, at Pico de Veleta (Spain), to
observe the mm-wave emission of $^{12}$CO and $^{13}$CO, $J$ = 1--0 and
$J$ = 2--1, and of SiO \jcc, in the symbiotic stars R Aqr, CH Cyg, and
HM Sge. As calibration standards, we also observed the sources
IRC\,+10216, CRL\,2688, and NGC\,7027, whose CO emission is strong and
well characterized.

The observations were performed in two observing sessions, in May 2008
and May 2009.  During the first session we used the A100 and B100
receivers for the 3\,mm band, and the A230 and B230 for the 1\,mm band,
to observe four lines simultaneously. The data were recorded using the
1\,MHz filterbank and the VESPA autocorrelator.  For the second
session, we used the new EMIR\footnote{Eight MIxer Receiver} receiver,
observing simultaneously in the E090 and E230 bands (3\,mm and 1\,mm)
in dual polarization mode. In this run, only the $^{12}$CO lines were
observed.  The data was recorded using the WILMA and VESPA
autocorrelators.

The spatial resolution of the observations at 3\,mm is 23$''$, and
12$''$ at 1\,mm wavelength. Frequent pointing measurements were
performed to measure and correct pointing errors; typically, errors no
higher than 3$''$ were found, which practically have no effects on the
calibration. The observations were done by wobbling the subreflector by
2$'$ at a 0.5\,Hz rate. This method is known to provide very stable and
flat spectral baselines

The atmospheric conditions during the observations were good. The
average zenith opacity during both observing runs was $\sim$ 0.2,
slightly better at 110 GHz and slightly worse at 115 GHz. 

The data presented here has been calibrated in units of
(Rayleigh-Jeans-equivalent) main-beam temperature, corrected for the
atmospheric attenuation, $T_{\rm mb}$, using the standard chopper
wheel method. Calibration scans (observation of the hot and cold loads,
and of the blank sky) have been performed typically every 15--20
minutes. The temperature scale is set by observing hot and cold loads
at ambient and liquid nitrogen temperatures. Correction for the antenna
coupling to the sky and other looses have been done using the latest
values for these parameters measured at the telescope. The accounting
for the sky attenuation is computed from the values of a weather
station, the measurement of the sky emissivity, and a numerical model
for the atmosphere at Pico de Veleta.  Finally, we checked that the
calibration of the different observations was compatible. We also
compared the intensities of IRC\,+10216, CRL\,2688, and NGC\,7027 with
previous observations to check calibration uncertainties; we took
observational data in \cite{bujetal01}, but taking into account that
recent measurements suggest that the 1\,mm data in that paper seem to
be overcalibrated by about a 20\%. The corrections applied to the
calibration of the different observations were always moderate, not
larger than $\sim$ 20\%, which can be considered as a measure of the
absolute calibration uncertainty.

All the data were averaged and rebinned, to get an adequate velocity
resolution. Baselines of degree 1 or 2 were subtracted. We note that in
the EMIR observations a source of noise at intermediate frequencies was
detected, in such a way that averaging the WILMA and VESPA data a
slightly smaller noise was obtained. (Averages of observations obtained
simultaneously with different spectrometers is not a standard procedure
in radioastronomy, because the noise from low-frequency parts of the
detection chain is usually negligible and this procedure does not help
to improve the S/N ratio.) In any case the differences are small; as an
example, the spectra shown in Fig.\ 4 (to be compared to that shown in
Fig.\ 1) was obtained after averaging both spectrometers.

\begin{figure}
  \vspace{0.2cm}
\rotatebox{270}{\resizebox{7.5cm}{!}{ 
\includegraphics{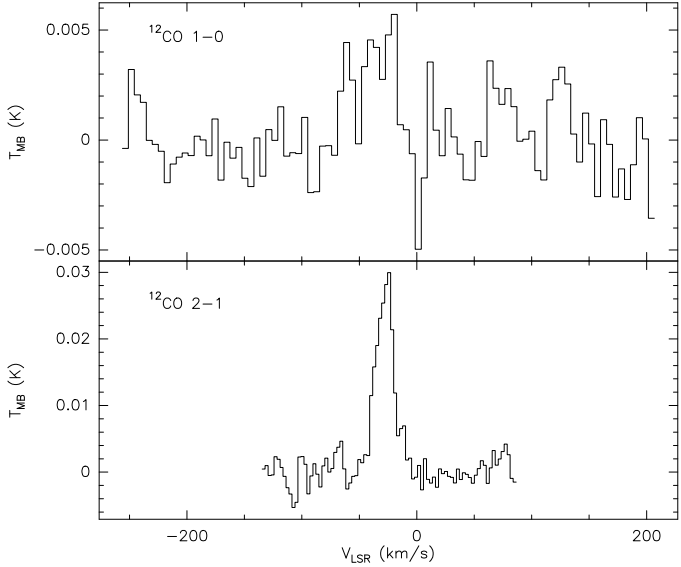}
}}
\caption{\doce\ lines detected in R Aqr.}
\end{figure}

\begin{figure}
  \vspace{0.2cm}
\rotatebox{270}{\resizebox{7.5cm}{!}{ 
\includegraphics{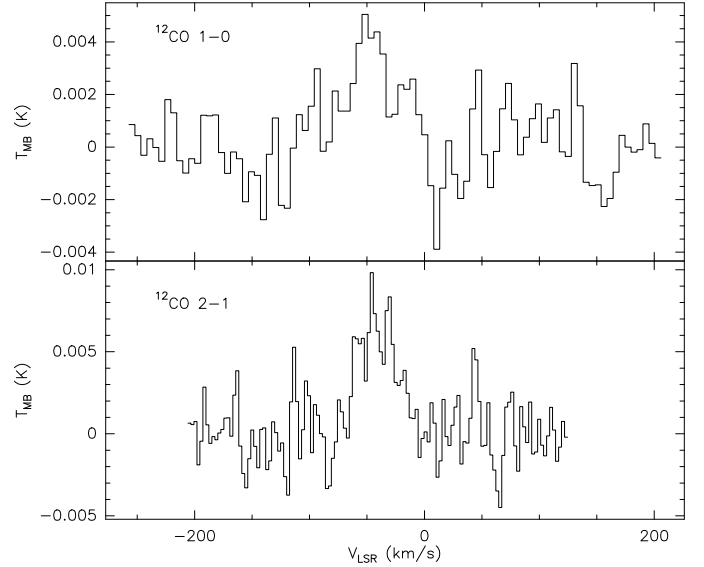}
}}
\caption{\doce\ lines detected in CH Cyg.}
\end{figure}

\section{Line formation in inner circumstellar shells around symbiotic
  stellar systems}

\subsection{Simple predictions of molecular line intensity}

As discussed in Sect.\ 1, the properties of molecular lines from
SSs, particularly their weak emission, are thought to be due to effects
of the companion on the outer CSE, by means of
photodissociation or of dynamical disruption. If the detected CO
emission comes from regions closer than the separation between both
stellar components (several AU, 1--2 $\times$ 10$^{14}$ cm), it must
present very different properties than the usual circumstellar CO
lines, which come from regions as far as $\sim$ 10$^{17}$ cm.

The excitation of the low-$J$ CO transitions is in general very easy to
describe, the population of the levels only requires temperatures \gsim\
15 K and is thermalized (i.e. in LTE) for densities larger than 
10$^4$ cm$^{-3}$. Temperatures close to 1000 K and densities \gsim\
10$^8$ cm$^{-3}$ are expected in the innermost circumstellar layers
around AGB stars, say at a few stellar radii (Sect.\ 1). Even for
extended envelopes, these conditions are satisfied in practically all
the emitting region \citep[see discussion in e.g.][]{teyssier06}. CO
lines from regions at $\sim$ 10$^{14}$ cm are expected to be very weak
compared to those from standard envelopes, due to the large dilution
factor in that case.

We can perform simple, but accurate estimates of the emission of CO
low-$J$ lines from the inner CSE (the predictions of
our simple calculations will be checked later on by means of numerical
calculations). Due to the very high densities expected in these
regions, the populations of all the relevant CO rotational levels are
accurately thermalized. (It is easy to show that the population of the
vibrationally excited levels is negligible, since the vibrational
de-excitations are very probable and the stellar IR radiation is
strongly diluted a soon as we are placed at a few stellar radii.)
Accordingly, the population of each $J$-level, $n(J)$, is given by:
\begin{equation}
n(J) =  g_{j} \,x(J) = g_{j} \,x(0) e^{-E_{j}/k T_{\rm k}} ~~,
\end{equation}
where $x$ is the population per magnetic sublevel, $g$ is the
statistical weight (in our case, $g_{j}$ = $2J+1$), $E_{j}$ is
the energy of level $J$, and $T_{\rm k}$ is the kinetic
temperature. Other symbols in our formulae ($k$, $h$, ...) have their
usual meanings.

In terms of the total density, $n$, and the CO relative abundance,
$X$(CO): 
\begin{equation}
x(J) = n X e^{-E_{j}/k T_{\rm k}} / F(T_{\rm k}) ~~.
\end{equation}
$F(T_{\rm k})$, the partition function, is accurately given in our case
(a linear thermalized molecule, $T_{\rm k}$ $\gg$ $B$) by $F(T_{\rm
k})$ = $T_{\rm k}/B$, $B$ being the rotational constant in temperature
units (for CO, $Bh/k$ = 2.75 K). We recall that, in our case, for the
CO rotational line $J$ $\rightarrow$ $J - 1$ and properly adapting the
units of $B$: $E_{j}$ = $B J(J+1)$ and $\nu_{j}$ = $2 B J$.

With these expressions, we can easily write the representative
absorption coefficient and optical depth, $\tau$, in a medium with
characteristic kinetic temperature, density, CO abundance, velocity
dispersion ($\Delta V$) and size ($L$). We must keep in mind that in
our case the line profile is given by the velocity dispersion within
the source. For transition $J$
$\rightarrow$ $J - 1$: 
\begin{equation}
\tau(J) = n X \frac{1}{F(T_{\rm k})} \frac{c^3}{8 \pi \nu_{j}^3}
A_{j} g_{j} [ x(J-1) - x(J) ] \frac{L}{\Delta V}~~.
\end{equation}
The Einstein coefficient in this linear, simple molecule is $A_{j}$ =
$\frac{64 \pi^4 \nu{_j}^3}{3 h c^3} \mu^2 \frac{J}{g_{j}}$, $\mu$ being
the permanent dipole moment that in CO is particularly low.  Note that
here $\tau$ is the opacity at the central frequency of the line, $\nu$,
corresponding to the systemic velocity of the source $V_{\rm sys}$. For
nearby frequencies, $\nu \pm \Delta\nu/2$ ($V_{\rm sys} \pm \Delta
V/2$), $\tau$ should vary depending on the velocity distribution within
the source. (Of course, $\tau$ decreases strongly far from the
resonance frequency.)  We will see below the dependence of $\tau$ on
the observed velocity (i.e.\ on the observed Doppler-shifted frequency)
for the velocity fields expected in our sources, around their systemic
velocities.

The dependence of $\tau$ on $T_{\rm k}$ and $J$ is interesting. Since, in our
case (and in many cases for CO low-$J$ transitions), we can assume that
the level populations are thermalized and $T_{\rm k}$ $\gg$ $B$,
\begin{equation}
\tau(J)\sim C\, Xn\, \frac{L}{\Delta V}\, B^2 \mu^2\, \frac{J^2}{T_{\rm
k}^2} ~~,
\end{equation}
where the constant $C$ only includes usual mathematical and physical
constants.  Accordingly, for different values of the temperature,
\begin{equation}
\tau(J) ~~\aprop~~  1/T_{\rm k}^2 ~~,
\end{equation}
and, for different transitions (and the same $T_{\rm k}$), 
\begin{equation}
\tau ~~\aprop~~  J^2 ~~.
\end{equation}

The characteristic brightness of a source (neglecting the cosmic
background, whose temperature is much smaller than $T_{\rm k}$ in our
case) is
\begin{equation}
B(\nu) \sim  S  (1 -  e^{-\tau(\nu)}) ~~. 
\end{equation}
Where $S$, the source function, depends on the frequency much less
sharply than the line profile:
\begin{equation}
S = \frac{2 h \nu^3}{c^2}\frac{x(J)}{x(J-1)-x(J)} ~~.
\end{equation}
Since we are assuming that the populations are thermalized (eqs.\ 1, 2)
and, again, that $T_{\rm k}$ $\gg$ $B$, we can safely apply the
Rayleigh-Jeans approximation and write the brightness temperature as:
\begin{equation}
T_{\rm b}(\nu) \sim T_{\rm k} (1 -  e^{-\tau(\nu)}) ~~.
\end{equation}
Leading to the usual equations for the optically thick case,  
\begin{equation}
T_{\rm b} \sim T_{\rm k} ~~,
\end{equation}
and for the optically thin case,
\begin{equation}
T_{\rm b}(\nu) \sim T_{\rm k} \tau(\nu) ~~, ~~T_{\rm b}(V) \sim T_{\rm
k} \tau(V) ~~.
\end{equation}
Note that equation 10 holds for the range of velocities around the
systemic one for which the opacity is still higher than 1, and that
$T_{\rm b}$ in eq.\ 11 keeps the velocity dependence of $\tau$. We also note
that, if we allow the amount of emitting molecules to vary, the
optically thick case always represents the upper limit to the source
emission. 

From these equations we can deduce also the dependence of $T_{\rm b}$
on $T_{\rm k}$, taking in particular into account the dependence of
$\tau$ on $T_{\rm k}$ (see eq.\ 4): $T_{\rm b}$ $\propto$ $T_{\rm k}$
in the optically thick case, and $T_{\rm b}$ $\propto$ 1/$T_{\rm k}$ in
the optically thin case.

Finally, for a source with a typical size $L$, in arcsecond units,
observed with a telescope beam with typical half-power beam width equal
to $W_{\rm b}$, also in arcseconds, 
the main beam temperature $T_{\rm mb}$ (over the cosmic background)
will be:
\begin{equation}
T_{\rm mb} \sim T_{\rm b} (L/W_{\rm b})^2 ~~.
\end{equation}
[We are assuming that the typical source size $L$ represents the source
observable diameter at half-maximum, and that the dilution factor
within the beam, $(L/W_{\rm b})^2$, is much smaller than one, since the
source is in our case unresolved.]

\subsection{Line shapes: optically thin emission}

The typical profiles of molecular lines from CSEs 
around standard AGB stars, very extended and expanding at supersonic
almost constant velocities, are well known. Parabolic, flat-topped or
two-horn profiles are expected depending on whether the lines are
optically thin or thick and on whether the source is spatially resolved
by the telescope \citep[e.g.][]{olofsson82}.

The profile shapes are of course modified by the presence of
significant local velocity dispersions, which yield profiles that are
in some way the convolution of the above prototypes with the local
distribution (often assumed to be described by a Gaussian function). A
discussion on the line formation when the local velocity dispersion is
not negligible requires detailed computations. Therefore, we will assume
in this Section that the turbulence velocity is much smaller than the
macroscopic velocity field (which is very often the case in CSEs
around AGB stars, with expansion velocities
around 10 \kms\ and local dispersions $\sim$ 1 \kms).

In the very inner circumstellar shells we do not expect such constant
outwards velocities. When significant acceleration is still present,
simple sharply-peaked profiles are expected; see general discussion and
comparison with standard profiles in \citet{bujetal89}.

Simple considerations can yield analytical insight into the expected
profiles from these regions. We recall that, in our sources, we can
assume spherical symmetry and isotropic expansion, thermalized
populations, $T_{\rm k}$ $\gg$ $B$, unresolved sources, and a constant
abundance.

In the optically thin case, the velocity-integrated emission in a
certain CO line of a certain mass of gas $M$ at a temperature $T_{\rm
k}$ is just proportional to $M/T_{\rm k}$ (from eqs.\ 3 to
12). Therefore, the brightness temperature emitted by a geometrically
thin shell (radius: $r$) integrated over observable velocities, $V_{\rm
obs}$ (projected on a line of sight), would be
\begin{equation}
\int T_{\rm mb}(r,V_{\rm obs}) ~~{\rm d}V_{\rm obs} \propto 4 \pi r^2
n(r)/T_{\rm k} ~~.
\end{equation}
Some obvious geometrical considerations show that the brightness,
$T_{\rm mb}(r,V_{\rm obs})$, from the thin shell is independent of
$V_{\rm obs}$, provided that $V_{\rm obs}$ differs from the systemic
velocity, $V_{\rm sys}$, by less than plus/minus the expansion velocity
of the shell, $V(r)$. Since each shell emits at velocities between
$V_{\rm sys}$ -- $V$ and $V_{\rm sys}$ + $V$, we can write: $T_{\rm
mb}(r,V_{\rm obs})$ $\propto$ $4 \pi r^2 n(r)/T_{\rm k}/V(r)$.

If we assume a constant mass-loss rate, \mloss, $n(r)$ $\propto$
\mloss$/r^2/V(r)$, and 
\begin{equation}
T_{\rm mb}(r,V_{\rm obs}) \propto \mloss/T_{\rm k}/V^2(r) ~~.
\end{equation}
The total emission of the optically thin envelope, between its inner
and outer radii, $R_{\rm i}$ and
$R_{\rm o}$, is therefore:
\begin{equation}
T_{\rm mb}(V_{\rm obs}) ~\propto~ \mloss \int 
1/T_{\rm k}(r)/V^2(r) ~~{\rm d}r ~~.
\end{equation}
Where the integral extends only to shells such that $V_{\rm sys}$ --
$V(r)$ $\leq$ $V_{\rm obs}$ $\leq$ $V_{\rm sys}$ + $V(r)$, the only
ones that can emit at $V_{\rm obs}$.  When the expansion velocity is
constant, the resulting intensity does not depend on $V_{\rm obs}$
within $V_{\rm sys}$ $\pm$ $V$, being zero otherwise; the profiles
are therefore flat-topped, one of the cases discussed
by \cite{olofsson82}.

If we assume, for instance, a constant increase of $V$ with $r$, $V(r)$
= $V(R_{\rm o}) r / R_{\rm o}$, and constant $T_{\rm k}$, the variation
of $T_{\rm mb}$ with $V_{\rm obs}$ from eq.\ 15 becomes:
\begin{equation}
T_{\rm mb}(V_{\rm obs}) \propto \frac{V(R_{\rm o})}{|V_{\rm obs} -
V_{\rm sys}|} - 1 ~~
\end{equation}
Except for truncation for central velocities such that $|V_{\rm obs} -
V_{\rm sys}|$ $<$ $V(R_{\rm i})$, for which the intensity remains
constant. The line shape then shows a central maximum, which can be
quite sharp because usually $R_{\rm i}$ is much smaller than $R_{\rm
o}$.

If $T_{\rm k}$ varies like $1/r$, the solution is also sharply peaked:
\begin{equation}
T_{\rm mb}(V_{\rm obs}) \propto {\rm ln} \frac{V(R_{\rm o})}{|V_{\rm
obs} - V_{\rm sys}|}
\end{equation}
for $|V_{\rm obs} - V_{\rm sys}|$
$\geq$ $V(R_{\rm i})$, and constant for smaller values of $|V_{\rm obs} -
V_{\rm sys}|$.

Finally, we note that the conversion of the proportionality relations
given here into true equations (i.e.\ the substitution of $\propto$ for
=) is straightforward using the complete expression for 
$\tau$ as given in eq.\ 3. 

\subsection{Line shapes: optically thick emission}
 
As in Sect.\ 3.2, our model CSE presents spherical symmetry and
isotropic expansion.  Let us assume now that the emission from a
geometrically thin shell within a velocity range, $\Delta V_{\rm obs}$,
centered around the systemic velocity $V_{\rm sys}$, is optically
thick. In general, $\Delta V_{\rm obs}$/2 can be assumed to be equal to
the maximum expansion velocity in the CSE.  The corresponding main-beam
temperature is then just proportional to the kinetic temperature,
$T_{\rm k}$, and to the solid angle from which emission at this
velocity is emitted (Sect.\ 3.1). Let's also assume that the expansion
velocity $V(r)$ is constant or monotonically increasing with the
distance to the center, in order to simplify the formulae, and again a
small local velocity dispersion. Also for the sake of simplicity in
formulae, we will assume that $V_{\rm sys}$ = 0; in other words, our
observed velocity $V_{\rm obs}$ is considered in this subsection to be
measured with respect to the characteristic velocity of the source.

The total emission in main-beam temperature units of the circumstellar
envelope (with inner and outer radii $R_{\rm i}$ and $R_{\rm o}$),
within $\pm$ $\Delta V_{\rm obs}/2$ (i.e.\ within $V_{\rm sys}$ $\pm$
$\Delta V_{\rm obs}/2$), is then
\begin{equation}
T_{\rm mb}(V_{\rm obs}) = F_{d} 2 \pi \int^{R_{\rm o}}_{R_{v}} p
T_{\rm k}(r) ~~{\rm d}p ~~.
\end{equation}
Where $R_{v}$ is the minimum radius emitting at $V_{\rm obs}$, taking
into account projection on the line of sight [i.e. $R_{v}$ is the
minimum $r$ within [$R_{\rm i}$,$R_{\rm o}$] such that $|V_{\rm obs}|$
$\leq$ $V(r)$]. In this equation, $F_{d}$ is the inverse solid angle of
the telescope beam in length units for a given distance to the source,
$D$. For a Gaussian beam, $F_{d}$(cm) = $\frac{1}{W_{\rm b}({\rm
cm})^2\frac{\pi}{4\,{\rm ln}2}}$~, with $W_{\rm b}$(cm) = $W_{\rm
b}$(arcsec)\,$D$(pc)\,1.5\,10$^{13}$ and $W_{\rm b}$(arcsec) being the
half-maximum full width of the beam; $W_{\rm b}$(cm) is always assumed
here to be much larger than the outer radius $R_{\rm o}$(cm), as
expected in our case. Finally, $p$ is the varying impact parameter,
i.e. the radius in the plane of the sky of the differential ring that
is emitting at $V_{\rm obs}$ and is placed at given distance to the
center $r$. Due to the projection, $p$ satisfies:
\begin{equation}
p = r \sqrt{1 - V^2_{\rm obs}/V^2(r)} ~~.
\end{equation}

Therefore, the main-beam brightness temperature in the optically
thick case is given by:
\begin{eqnarray}
T_{\rm mb}(V_{\rm obs}) = F_{d} 2 \pi \int^{R_{\rm o}}_{R_{v}}
T_{\rm k}(r) r \sqrt{1 - V^2_{\rm obs}/V^2(r)} ~~\times \nonumber \\
\left[ \frac{V(r) - r V'(r)}{V^2(r)} \sqrt{V^2(r)-V^2_{\rm obs}} +
\frac{r}{V(r)} \frac{V(r)V'(r)}{\sqrt{V^2(r) - V^2_{\rm obs}}} \right]
~{\rm d}r ~.
\end{eqnarray}

These equations systematically lead to profiles with a central peak,
but less sharp than in the optically thin case. For instance, if the
expansion velocity is constant, $V(r)$ = $V_{\rm exp}$, 
we get the well known parabolic profiles \citep{olofsson82}:
\begin{equation}
T_{\rm mb}(V_{\rm obs}) = F_{d} 2 \pi \left( 1 - \frac{V^2_{\rm
obs}}{V^2_{\rm exp}} \right) \int^{R_{\rm o}}_{R_{\rm i}} r T_{\rm
k}(r) ~~{\rm d}r ~~.
\end{equation}

In the case of a constant velocity gradient, $V(r)$ = $V(R_{\rm o}) r /
R_{\rm o}$, which probably applies in our sources,  
the solution is also simple:
\begin{equation}
T_{\rm mb}(V_{\rm obs}) = F_{d} 2 \pi \int^{R_{\rm o}}_{R_{v}}
r T_{\rm k}(r) ~{\rm d}r ~~.
\end{equation}
For the assumed velocity field, $R_{v}$ = $|V_{\rm obs}| R_{\rm o}
/ V(R_{\rm o})$; except when $R_{\rm i}$ $>$ $|V_{\rm obs}| R_{\rm o} /
V(R_{\rm o})$, then $R_{v}$ = $R_{\rm i}$. This property leads
systematically to a truncation of the central parts of the profile,
$|V_{\rm obs}|$ $<$ $V(R_{\rm i})$, for which $T_{\rm mb}(V_{\rm obs})$
is constant and equal to $T_{\rm mb}$($V_{\rm obs}$=$V(R_{\rm i})$). We
note that, however, $R_{\rm i}$ is usually much smaller than $R_{\rm
o}$ in circumstellar envelopes around evolved stars, and then the
effects of truncation are negligible.

For constant $T_{\rm k}$, 
parabolic profiles are again predicted: $T_{\rm mb}(V_{\rm obs})$ =
$F_{d} R_{\rm o}^2 \pi T_{\rm k}$ $[1 - V^2_{\rm obs}/V^2(R_{\rm
o})]$, assuming very small $R_{\rm i}$ (otherwise the profiles are
parabolic except for truncation in the central velocities).

It is, however, more realistic to assume that the temperature decreases
outwards, which leads to more sharply peaked shapes. If, for instance,
$T_{\rm k} = T_{\rm k}(R_{\rm o}) R_{\rm o} / r$ [and, we recall,
$V(r)$ = $V(R_{\rm o}) r / R_{\rm o}$], we get from eq.\ 22 sharp
profiles that are exactly triangular if $R_{\rm i}$ $\ll$ $R_{\rm o}$:
\begin{equation}
T_{\rm mb}(V_{\rm obs}) = F_{d} 2 \pi T_{\rm k}(R_{\rm o}) R^2_{\rm
o} \left[ 1 - \frac{|V_{\rm obs}|}{V(R_{\rm o})} \right] ~~.
\end{equation}

When the temperature decreases linearly from $R_{\rm i}$ to $R_{\rm
o}$ and again for $R_{\rm i}$ $\ll$ $R_{\rm o}$, eq.\ 22 gives:
\begin{eqnarray}
T_{\rm mb}(V_{\rm obs}) = F_{d} \pi R^2_{\rm
o} T_{\rm k}(R_{\rm i}) [1 -  V^2_{\rm obs}/V^2(R_{\rm o})] ~~-
\nonumber \\ 
\frac{2}{3}
F_{d} \pi [T_{\rm k}(R_{\rm i}) - T_{\rm k}(R_{\rm o})] [1 -  |V_{\rm
    obs}|^3/V^3(R_{\rm o})] ~~.
\end{eqnarray}
In this case, the profile is more sharply peaked than the parabolic
profile obtained for constant $T_{\rm k}$, but without the
discontinuity in its slope at $V_{\rm obs}$ = 0 characteristic of
triangular profiles. Truncation also appears for $|V_{\rm obs}|$ $<$
$V(R_{\rm i})$ and would be noticeable in the unexpected case that
$R_{\rm i}$ is not much smaller than $R_{\rm o}$.

These equations are particularly interesting for us, because probably
the kinetics of the circumstellar shells in our sources shows a 
significant velocity gradient and our \doce\ lines are optically
thick. We therefore expect more or less triangular profiles with a
central peak. As we will see below, other phenomena should also be
taken into account to explain the asymmetry in the observed profiles.

\subsubsection{Selfabsorption in optically thick envelopes}

In discussion in Sects.\ 3.2 and 3.3, we have assumed that the local
velocity dispersion, presumably due to turbulent movements, is much
smaller than the macroscopic velocities. A variety of line profiles
were found, but, in particular, in all cases the profiles were
symmetric around the systemic velocity $V_ {\rm sys}$. If the local
velocity dispersion is not negligible (and the temperature is not
constant), selfabsorption must be taken into account. In the most usual
case, in which the temperature decreases outwards, selfabsorption is
important for the most negative velocities (gas approaching the
observer), since then outer cool regions absorb the emission of hotter,
inner regions. This is not the case of the positive velocities, in
which the hotter regions would be closer to the observer, its emission
dominating the observed line. The result is an asymmetric profile in
which the maximum is shifted to positive velocities, while keeping the
total emission range.

Unfortunately, these effects are too complex to be discussed by means
of analytical, simple formulae, as those given before, and numerical
calculations must be performed. In Sect.\ 3.4 we will present a
numerical model that takes into account selfabsorption, within an
accurate description of radiative transfer in expanding envelopes. We
will also see in Sect.\ 4 how these phenomena are useful to better
understand our observations of SSs, particularly the asymmetric profile
of the \jdu\ line observed in R Aqr.

\subsection{Numerical model of line formation}

We have used a numerical model to simulate the line formation in our
case similar
to that described by \citet{teyssier06}, adapted to the high
densities and temperatures expected in the inner circumstellar
shells. See more details on the numerical code in that paper.

In the calculation of the level population, we have taken into account
collisional and radiative excitation, considering a high number of
rotational levels in two vibrational states. Our calculations confirm
that, as shown in Sect.\ 3.1, the population of all relevant levels is
very accurately thermalized due to the expected high
densities. Therefore, our results practically do not depend on the
details of the treatment of the excitation.
The radiative transfer equation is accurately solved taking into
account a local turbulence velocity and the macroscopic velocity
field. We so calculate the brightness in the direction to the observer
along a number of lines of sight and for a number of projected
velocities. These results are convolved with the beam shape assumed for
the observed lines, obtaining a Rayleigh-Jeans--equivalent main-beam
temperature for a number of {\em LSR} velocities, directly comparable
to the observed profiles. Thanks to the small extent of the source,
this convolution practically does not depend on the exact beam shape; 
the beam width is therefore the only relevant parameter in the
calculation.

In this model, we are assuming that the emitting region is spherical
and expands isotropically. We also assume that the physical and
chemical conditions are in our case similar to those typical of inner
shells around standard AGB stars. This may be not true in symbiotic
systems, whose inner circumstellar layers could present an axis of
symmetry, anomalous density and velocity distributions, and other
peculiar properties. We think that our lack of knowledge on these
probably complex regions (from molecular line observations or from
other studies) prevents a more detailed modelling. We note that
observations of SiO maser emission in R Aqr
\citep[e.g.][]{cotton04,kamohara10,pardo04} show properties completely
similar to those found in standard AGB stars. VLBI maps show, in
particular, that the SiO maser in general occupy a ring-like region
(with radius $\sim$ 5 10$^{13}$ cm) that is comparable to those typical
in AGB stars. Kamohara et al.\ found that the spatial distribution
tends, nevertheless, to show a certain axial symmetry. The departures
from spherical symmetry do not always show the same pattern and are in
many epochs quite small, but the axis orientation has remained stable
at a position angle of about --10$^\circ$ during more than 10
years. The SiO data therefore indicate that the inner shells around R
Aqr are more or less spherical and, because SiO masers are excited
normally, must show physical conditions similar to those usually
present in AGB stars; at least up to a distance of about 5 10$^{13}$
cm, i.e. in regions not much smaller that the total region assumed to
be responsible for the CO lines in our model.

\section{Results}

We have detected \doce\ \jdu\ and \juc\ emission from the symbiotic
stellar systems (SSs) R Aqr and CH Cyg (the detection of \juc\ in R Aqr
being tentative). We also present limits of the emission of \doce\
\jdu\ and \juc\ in another SS, HM\,Sge, and of the emission of \trece\
\jdu\ and \juc\ and SiO $J$=5--4 in R Aqr. A summary of our
observational results can be seen in Table 1 and the detected lines are
shown in Figs.\ 1, 2. The interpretation of these data can be done by
means of the relatively simple formulation presented in Sects.\ 3.1 -
3.3; but, as we will see, some observational features require numerical
calculations, using the code presented in Sect.\ 3.4. We will see that
the observations are reasonably well explained by those predictions,
and other cases that do not seem compatible with our data, like the
narrow two-horn profiles from rotating disks, will not be considered.

\begin{table}[bthp]
\caption{Summary of observational results: 
peak main-beam temperature or 3$\sigma$ limit,
  1$\sigma$ noise, and spectral resolution with which the noise was
  calculated.}
\vspace{0.1cm}
\begin{center}
\begin{tabular}{|l|c|c|c|}
\hline
 & & &  \\
source & line &  peak T$_{\rm mb}$ or 3$\sigma$ limit  & 1$\sigma$ 
 noise ~~(res.) \\
       &      &              mK                   & mK   \\
\hline 
 & & & \\
R Aqr    & $^{12}$CO 2--1 & 30            & 2 ~~(2.6 \kms) \\
         & $^{12}$CO 1--0 & $\sim$ 6      & 2 ~~(5.2 \kms) \\
         & $^{13}$CO 2--1 & $<$ 18 & 6 ~~(2.7 \kms) \\
         & $^{13}$CO 1--0 & $<$ 18 & 6 ~~(2.6 \kms) \\
         & SiO 5--4       & $<$ 19 & 6 ~~(2.8 \kms) \\
\hline
 & & & \\
CH Cyg   & $^{12}$CO 2--1 & 9      & 2 ~~(2.6 \kms) \\
         & $^{12}$CO 1--0 & $\sim$ 4.5      & 1.4 ~~(5.4 \kms) \\
\hline
 & & & \\
HM Sge   & $^{12}$CO 2--1 & $<$ 19 & 6.5 ~~(2.6 \kms) \\
         & $^{12}$CO 1--0 & $<$ 26 & 9 ~~(2.6 \kms) \\
\hline
\end{tabular}
\end{center}
\end{table}

\subsection{CO lines in R Aqr}

Let us apply the simple formulae in Sect.\ 3 to our observations of CO
in R Aqr. We will first assume that the radius of the emitting region
is smaller than the component separation, i.e.\ that $L$ $\sim$ 0\farcs
09 (3 $\times$ 10$^{14}$ cm at a distance of 214 pc). Then, the upper
limit to the antenna temperature is given by the optically thick case
emission: $T_{\rm mb}$(\doce \jdu) $\sim$ $T_{\rm k} \times 6\,
10^{-5}$ K. The observed intensity in R Aqr implies a typical kinetic
temperature of about 500 K; slightly higher temperatures will be
derived if the line opacity is not extremely large, which, as we will
argue below, is probably the case. From eq.\ 3, Sect.\ 3.1, and
assuming a CO abundance of about 5 $\times$ 10$^{-4}$, typical in CSEs
around O-rich Mira-type stars, we deduce that the assumption of
optically thick \doce\ \jdu\ emission implies densities \gsim\ 5 10$^8$
cm$^{-3}$. As we have discussed in Sect.\ 1, both deduced values of the
temperature and density agree with current expectations for the
innermost shells around AGB stars.

We have discussed in Sect.\ 3.1, eqs.\ 6, 10, 11, and 12, the expected
line ratios in our case.  In the completely opaque case, the intensity
ratio of the \jdu\ and \juc\ lines must be $T_{\rm mb}$(\jdu)/$T_{\rm
mb}$(\juc) $\sim$ $W_{\rm b}^2$(\juc)/$W_{\rm b}^2$(\jdu) $\sim$
3.5. In the fully optically thin case, $T_{\rm mb}$(2--1)/$T_{\rm
mb}$(1--0) $\sim$ $\tau$(2--1)/$\tau$(1--0) $\times$ $W_{\rm
b}^2$(1-0)/$W_{\rm b}^2$(2--1) = 4\,$W_{\rm b}^2$(1--0)/$W_{\rm
b}^2$(2--1) $\sim$ 14. The line $T_{\rm mb}$ ratio is $\sim$ 6 (if we
believe that our CO \juc\ line is detected, \gsim\ 6 otherwise; in any
case, the uncertainty is very large). Therefore, our observations
suggest that the CO \juc\ line in this source is moderately optically
thin, typical values of $\tau$(\juc) $\sim$ 0.8 -- 1 are enough to
explain the (tentatively) measured line ratio. Since the typical $\tau$
ratio is $\tau$(\jdu)/$\tau$(\juc) $\sim$ 4, the \jdu\ line would be in
any case opaque. These values of $\tau$(\juc) imply densities $\sim$
10$^9$ cm$^{-3}$ (from formulae in Sect.\ 3.1), very reasonable in this
context. {The total mass in the emitting shell is therefore deduced
to be $\sim$ 2.5 $\times$ 10$^{-5}$ \ms. }

Other possibilities that could explain the observations are much less
probable. A significantly larger extent would imply temperature almost
decreasing with 1/$L^2$, to keep the observed intensity of the \jdu\
line. For instance, $L$ $\sim$ 10$^{15}$ cm would imply an average 
temperature of $\sim$ 45 K within the emitting region, surprisingly
low for shells around Mira-type stars (mostly taken into account 
the presence of a hot companion).
Values of $L$ as small as
$\sim$ 10$^{14}$ cm would imply typical temperatures of about 5000 K,
even higher than the surface temperature of the cool star.

These intensity predictions, deduced from a simple but robust line
formation model, have been fully confirmed by computations performed
with the model presented in Sect.\ 3.4. (Calculations performed
with this code will be discussed in more detail below.) 

The shape of our \doce\ \jdu\ profile is clearly different from the
wide profiles observed in the extended envelopes around standard AGB
stars (parabolic, flat-topped or showing two horns). The single-peak, 
relatively sharp profile is compatible with the expected profile for
the compact, very small shells expected in SSs (Sects.\ 3.2, 3.3), in
which a significant acceleration of the gas is expected to be present. 

Therefore, both the intensity of the CO \jdu\ and \juc\ lines and the
shape of \jdu\ are compatible with our expectations for the CO
emission in circumstellar regions closer than about 2 $\times$ 10$^{14}$
cm. 

To better understand the properties of the emitting region and check
the above conclusions, we have tried to fit our CO observations of R
Aqr with the predictions of the numerical line formation model
presented in Sect.\ 3.4. We must keep in mind that the source modeling
and the determination of the properties of this shell can be only a
first step in the study of these layers, because of the lack of
information still existing on them, including, in particular, the poor
data on molecular emission. {Moreover, in the inner circumstellar
shells around the cool component of an interacting binary system, we
can expect complex structures and velocity fields, perhaps difficult to
compare with the smooth, continuous ejection of material by standard
AGB stars, and yielding line shapes with features that cannot be
accounted for by our simple shell model.} The best we can do is to find
a set of characteristic properties of the emitting region, namely,
extent, velocity field and physical conditions, that yield predicted
line profiles compatible with the observations.

\begin{figure}
\vspace{-0.1cm}
\hspace{-0.5cm}
\rotatebox{270}{\resizebox{6.5cm}{!}{ 
\includegraphics{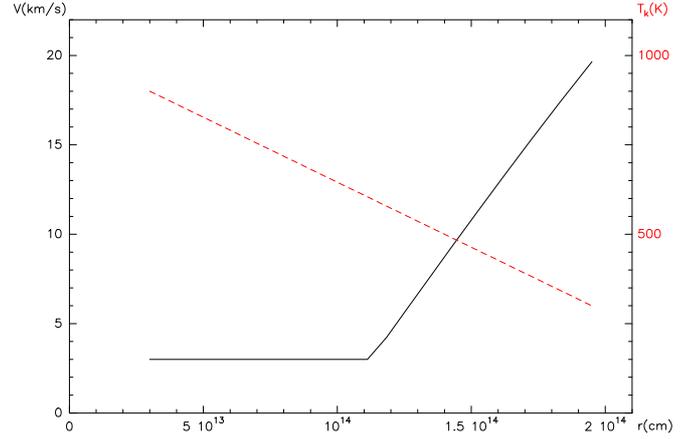}
}}
\vspace{-0.4cm}
\caption{Distributions of the expansion velocity (left scale) and kinetic
  temperature (red dash line, right scale) deduced from our simple
  modelization of the CO-rich, inner shells around R Aqr.}
\end{figure}

The envelope extent is defined by its inner and outer radii, $R_{\rm
i}$, $R_{\rm o}$. We have imposed a small value of $R_{\rm i}$,
slightly larger than the stellar radius, and a value of $R_{\rm o}$
equal to 2 $\times$ 10$^{14}$ cm, to be in agreement with general
considerations discussed above.  We have assumed that the velocity,
$V(r)$, increases from a certain point in the inner envelope (supposed
to be related to dust formation).
We assume a turbulence velocity comparable to the minimum expansion
velocity, $V(R_{\rm i})$. The temperature is assumed to vary linearly
in the considered region, between $T_{\rm k}(R_{\rm i})$ and $T_{\rm
k}(R_{\rm o})$, and the density varies assuming a constant mass-loss
rate, \mloss, i.e.: $n(r)$ = $\frac{\mloss}{4\pi r^2 V(r)}$. For both
the kinetic temperature and the expansion velocity, only linear
functions are considered, in view of our poor knowledge on these
parameters in our case. The CO abundance, $X$(CO), is assumed to be
constant and equal to 5 $\times$ 10$^{-4}$, a typical value in CSEs
around O-rich AGB stars. Note that, since the level population is
practically thermalized, we can vary the value of $X$(CO) and obtain
practically the same results, provided that the density varies in the
opposite sense and the product $n$\,$\times$\,$X$ remains constant.

We have found a set of values for the above parameters that can explain
the observations. The temperature and velocity laws are given in Fig.\
3. We note that in the outer shells around standard AGB stars the
velocity gradient becomes very small and the expansion velocity
increases slowly up to a certain asymptotic value, but this regime
apparently is not reached in the compact shells around R Aqr we are
describing here. The best-fit value of the mass-loss rate is 9 $\times$
10$^{-6}$ \my. In Fig.\ 4 we show the predicted profile, superimposed
to the observed one.

Because of the poor observational data on \doce\ \juc, we have not
tried to fit the observed profiles, but we have checked that the
predicted intensity, $T_{\rm mb}$ $\sim$ 5 mK, is compatible with the
observation.

Note that the asymmetric line shape, with a slightly red-shifted
maximum, is well reproduced by our model, due to absorption of the
emission of inner shells by the cooler outer regions, only noticeable
at relatively negative velocities (Sect.\ 3.3.1).
The expansion velocity (a parameter almost directly given by the
observed profile width) is also quite high, compared to other AGB
shells, compatible with the idea that mass ejection is particularly
efficient in SSs (Sect.\ 1).

Given the lack of information about the emitting region and the
relatively simple model we are using, it is obvious that the values
deduced here for the different parameters are indicative and relatively
uncertain. Nevertheless, we think that the fact that we are able to
reasonably fit the observations shows that our basic assumptions about
the emitting region (size, temperature and velocity ranges, densities,
...) are essentially correct. This conclusion is supported by
observational results of SiO maser emission in this object, see Sect.\
3.4, which suggest that inner shells around R Aqr (not much smaller
than our CO emitting region) are more or less spherical and show
physical conditions similar to those typical in AGB stars. On the other
hand, moderately oblate density distributions have been proposed for
the inner circumstellar regions in SSs \citep{gav02,pods07}, as due to
gravitational focusing. From the existing data, we cannot discriminate
between these oblate distributions and more spherical ones. Since the
binary orbit in R Aqr is nearly edge-on, the assumption of such an
oblate structure would limit our requirements of relatively high column
densities to its equator. Therefore, the values of the mass-loss rates
for R Aqr would be, in this case, somewhat smaller that those given
above; we deduce smaller values of \mloss\ by a factor of $\sim$ 2. A
similar smaller value of the mass-loss rate can be obtained, in
general, if we allow the radius of the spherical shell to increase
moderately, for instance up to the separation between the stars, 2.3 --
2.5 10$^{14}$ cm. The data could be also reproduced assuming the source
to be somewhat smaller than in our standard model, the gas temperature
and the mass-loss rate being then deduced to be larger. (Important
variations of the source size would lead to unexpected properties of
the emitting gas, as mentioned before.) {Strong variations in the
total mass and mass-loss rate in the case of a significantly clumpy
medium are also improbable, provided that the total size does not vary
a lot, because the typical opacity and column density required to
explain the observations would not change.}

\begin{figure}
  \vspace{0.1cm}
\begin{center}
\rotatebox{270}{\resizebox{6.2cm}{!}{ 
\includegraphics{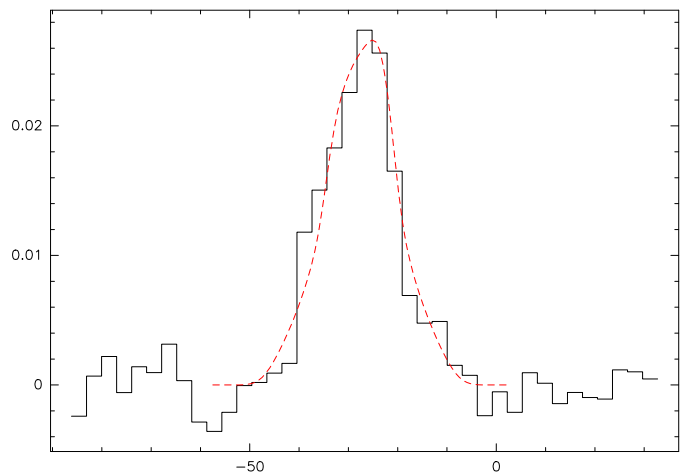}
}}
\caption{\doce\ \jdu\ profile predicted by our best-fit model of the
  inner shells around R Aqr, red dashed line, superimposed to our
  observed profile.}
\end{center}
\end{figure}

Published estimates of the mass-loss rates from the Mira component of R
Aqr (and in general for symbiotic Miras) are based either on radio
continuum observations or FIR IRAS data
\citep[e.g.][]{kenyon88,seaquist90}. The former method probes the
ionized portion of the wind, whereas the later measures the amount of
dust in the system.  In particular, \cite{seaquist90} estimated a
mass-loss rate from the Mira of 3.7 $\times$ 10$^{-8}$ \my\ from 6\,cm
radio flux, which would become 3 10$^{-8}$ \my\ for the distance value
adopted here and an expansion velocity $V_{\rm exp}$ $\sim$ 15 \kms
(note that there is a typing error in Table IV of Kenyon et al.\ 1988,
they should give the same value for the gas mass-loss rate as Seaquist
\& Taylor because both papers used the same data).

The IRAS data for symbiotic stars were analyzed by \cite{anandarao86},
\cite{anandarao88}, and \cite{kenyon88}.
The first group, based on analysis of the IRAS LRS data and the
broad-band IRAS photometry, found two dust shells in R Aqr: a hotter
inner shell with a temperature of about 800 K, a radius of about 6 AU
or $\sim$ 10$^{14}$ cm (assuming $D$ $\sim$ 214 pc), and a dust mass of
$\sim$ 2.5 $\times$ 10$^{-7}$ \ms, which may overlap with the CO
region, and an outer cold shell with about 87 K, a radius of about 123
AU, and a mass of $\sim$ 5.7 $\times$ 10$^{-4}$ \ms.  The hot inner
shell alone can account for the LRS fluxes, whereas the second outer
shell explains the far-IR fluxes, mainly in the 60 and 100 micron IRAS
bands. A similar complex dust structure was found in all of seven
studied symbiotic systems, and was interpreted as a signature of
multiple circumstellar shells due to a discontinuous mass distribution.
\cite{anandarao88} also suggested that the inner shells represent the
basic envelopes of the cool components, whereas the outer shells are
circumbinary components.  If the inner shells indeed result from the
present mass ejection by the Mira, this value of dust mass requires a
mass-loss rate of about 10$^{-5}$ \my\ (assuming gas/dust mass ratio
$\sim$ 100 and $V_{\rm exp}$ $\sim$ 15 \kms), in good agreement with
our estimate based on CO line emission.

On the other hand, the dust shell parameters derived from broad-band
IRAS photometry by \cite{kenyon88}, temperature $\sim$ 450 K and radius
of $\sim$ 70 AU, are significantly different from the above values, and
they indeed are incompatible with the LRS data. Their dust shell mass,
$\sim$ 1.6 $\times$ 10$^{-7}$ \ms\ (recalculated for our distance), is
similar to the mass of the inner shell proposed by Anandarao et al.;
but, due to the larger extent of the shell, the mass-loss rate is
$\sim$ 7 $\times$ 10$^{-7}$ \my\ (assuming gas/dust mass ratio $\sim$
100, and $V_{\rm exp}$ $\sim$ 15 \kms), an order of magnitude
lower than our CO-based estimate.  Finally, \cite{gromaetal09} derived
recently a mass-loss rate \mloss\ $\sim$ 10$^{-6}$ \my\ for R Aqr, from
the K-[12] color versus \mloss\ relation calibrated for normal AGB
stars, but its extrapolation to our case is uncertain.

In general, the comparison between the mass-loss rates derived from
dust and molecular line emissions in R Aqr is very uncertain, because
the interpretation of both molecular and FIR data under our extreme
conditions is not standard and because the dust-to-gas ratio may be
significantly different in shells around SSs and in standard AGB
stars. The dust content in the molecule-rich shells in symbiotic
systems may be lower than in CSEs around normal evolved stars, because
we only detect CO line emission from very inner shells in which grains
are not yet completely formed, as in fact suggested by the strong
velocity gradient deduced from our modelization. On the other hand, it
is clear than grains can survive much more easily to the radiation from
the hot companion, therefore we could detect dust emission from regions
in which molecules do not exist.

The comparison of results on ionized and molecular gas also needs some
discussion. The high-density gas is, in general, much harder to ionize
and photodissociate by the hot companion because of self-shielding. We
have mentioned that hydrodynamical numerical simulations show that
flows from the Mira in a SSs can be concentrated towards the orbital
plane, resulting in a large-scale density enhancement in the orbital
plane and low-density polar regions \citep[see e.g.][and references
therein]{gav02}. One can then expect the molecular material (and to a
lesser extent the dust) to be placed near the orbital plane, as well as
the presence of two ionized regions in the poles (and probably outside
the orbit).
The methods based on
observations of molecular gas and dust would refer to the denser
material, probably lying close to the orbital plane, whereas the radio
continuum data would probe the low-mass regions.
As a result, one should get higher mass values from molecular lines and
dust than from radio continuum data. Our rates from CO are about 2
orders of magnitude higher than published values based on radio data,
in agreement with these expectations.

\subsection{CO lines in CH Cyg}

The poor \doce\ profiles observed in CH Cyg prevent any detailed
fitting of the line shape. Both lines show roughly a central peak and
could be compatible with the emission of a region with significant
velocity gradient. We can derive some characteristics of the emitting
region from the line intensity. First of all, we can see that the
\jdu/\juc\ intensity ratio is relatively high, compatible with the
optically thick ratio, $\sim$ 3.5 (Sect.\ 3.1; see also discussion for
R Aqr in 4.1). We also note that the component separation in the object
is $\sim$ 9 AU, smaller than that of R Aqr.  Therefore we can assume
that both lines are optically thick and come from a very compact
region. From our discussion in Sect.\ 3.1, we can deduce that the
observed line intensities are compatible with an emitting region size
(typical diameter) of about 10 AU (typical radius $\sim$ 5 AU, somewhat
smaller than the component separation), and typical kinetic temperature
of about 800\,K. {We stress that, because of the lack of good CO
profiles in this source, the assumption that the emission comes from a
small shell in expansion is less well founded than for R Aqr; this
uncertainty and the low S/N ratio obviously result in less accurate
conclusions from the data.}

Note that the detected profiles are, within the uncertainties, quite
wide, suggesting that high expansion velocities are present in the
CO-rich shells, $\sim$ 25 \kms, even larger than those measured for R
Aqr. {The mass-loss rates must also be quite high, also larger than
for R Aqr, to take into account the larger opacities and velocity
dispersion (which enters both in the opacity, eq.\ 4, and in the
estimate of the shell lifetime)}; we deduce, following the
prescriptions in Sect.\ 3.1, \mloss\ \gsim\ 2 10$^{-5}$ \my.  We
suggest that the high values of the mass-loss rate and expansion
velocity are partially due to the fact that the CH Cyg system is
tighter than that of R Aqr.

{The high mass-loss rate in CH Cyg may be surprising in particular
because it is a semiregular (SR) variable, and it is well known that SR
variables in general show significantly lower mass-loss rates than
Mira-type variables. However, the case of CH Cyg is peculiar. CH Cyg
presents a complex variation pattern that probably includes two
periods of about 100 and 760 days, plus other long term variations,
with a high overall variability amplitude, about 3 mag in the visible;
see the light curve by the AAVSO and Miko\l ajewski et al. (1992).
Miko\l ajewski et al.\ indeed classify this source as SRa variable. The
basic properties and evolutionary status of the SR variables of type
SRa and SRb were comprehensively discussed by
\cite{kersch92,kersch94,kersch96}, who found that SRa stars appear as
intermediate objects between Miras and SRb variables in all aspects,
including periods, amplitudes, and mass-loss rates. Gromadzki et
al. (2007) found that most giants in symbiotic systems reveal more or
less regular pulsations with periods in the range 50-400 days. They
also concluded that the presence of such a variability can account for
the relatively high mass-loss rates usually found in symbiotic stars as
compared with single field giants (Sects. 1, 5).

It is also remarkable that semiregular variables with amplitudes \gsim\
2.5 mag (like W Hya, GY Aql, T Ari, etc) tend to present strong SiO
maser emission, comparable to that of standard Miras
\citep{alcolea90}. Therefore, the density and general physical
conditions in the inner circumstellar layers should not significantly
differ from those of Mira-type variables. These and other SR stars (W
Hya, GY Aql, RX Boo, EP Aqr, X Her, ...) show dense extended envelopes
well detected in CO. It is also well known that a number of SRs (EP
Aqr, RX Boo, X Her, RS Cnc, etc) show extended shells with a clear axis
of symmetry \citep[e.g.][]{nakashima05}, similar to those found in
Mira-type stars in binary systems, like $o$ Cet and V Hya
\citep[e.g.][]{kahane96}.

On the other hand, see Sects.\ 4.1, 5, it is obvious that the mass
ejection by the evolved component of a SS can be seriously affected by
the stellar interaction, and we have seen that mass-loss rates in such
systems tend to be larger than in isolated AGB stars (Sects.\ 1, 5). It
is then to be expected that the interacting nature of the CH Cyg system
strongly affects the structure, dynamics, and density of the inner
circumstellar layers, helping to understand the measured high amount of
gas in these regions.

The presence of a relatively high amount of mass in the inner
circumstellar shells of CH Cyg is confirmed by its FIR dust emission
and by the identification of a hot dust-shell, a few times larger than
the stellar photosphere, that significantly contributes to the total
NIR flux \citep{pedretti09}.  } The values derived here for the
mass-loss rate and typical temperature in the inner shells are quite
similar to those found by Taranova \& Shenavrin (2007) from analysis of
the dust FIR emission of recently ejected material. Our mass-loss rates
are also compatible with the total dust mass derived by Kenyon et al.\
(1988) and Hinkle et al.\ (2009), if we assume that dust emission comes
from inner shells not much larger than those we are detecting in CO
emission.

\subsection{Other molecular lines}

The other molecular observations in R Aqr, CH Cyg and HM Sge did not
yield detections. The upper limits obtained for \trece\ lines are
compatible with \doce/\trece\ abundance ratios larger than 10, as
usually found in similar objects. The nondetection of \doce\ to a limit
of $T_{\rm mb}(\jdu)$ $<$ 0.02 K in HM Sge, a source placed at more
than 1 kpc, is to be expected if the properties of the emitting region
are similar to those found for R Aqr and CH\,Cyg.

The nondetection of SiO \jcc\ thermal ($v$=0) emission in R Aqr is more
significant. This line is observed to be much weaker than \doce\
\jdu. Most silicon is expected to be in gas-phase SiO in these O-rich
regions in which dust grains are not formed, leading to relative
abundances of about 1 -- 5 $\times$ 10$^{-5}$. The Einstein A
coefficient of SiO \jsc\ is 5 $\times$ 10$^{-4}$ s$^{-1}$, almost 1000
times larger than that of \doce\ \jdu. Therefore, the SiO \jcc\ line
should be optically thick in R Aqr, if it comes from a region similar
to that of CO lines. However, SiO \jcc\ is more than 2 times weaker
than CO \jdu, so the size of the SiO emitting region in R Aqr should be
at least 2 times smaller than for CO (taking into account the increase
of temperatures in inner regions).
We therefore conclude that the SiO thermal emission in R Aqr only can
come from a relatively small region, probably not much larger than that
measured for SiO masers ($\sim$ 10$^{14}$ cm in diameter, Sect.\
1). This conclusion is interesting because SiO in standard AGB stars
must be very abundant in more extended layers, at least in regions in
which the grains are not completely formed, from which its intense
mm-wave emission comes. Indeed, the extent of the SiO thermal emission
in standard AGB stars has been found to be larger than 10$^{15}$ cm
\citep{lucas92}. The relatively small SiO emitting region in R Aqr
should therefore be due to that the molecular emission extent in this
source is limited by photodissociation due to UV emission from the hot
companion, rather than by dynamical distortion of the emitting region
(Sect.\ 1, 5), because that phenomenon affects significantly more the
SiO abundance than that of CO. We recall that CO is the most extended
molecule in CSEs because self-shielding yields less efficient
photodissociation for this molecule than for less abundant species, see
e.g.\ \cite{mamon88,willacy97}.

\section{Conclusions}

We present observations of the \doce\ and \trece\ \jdu\ and \juc\
transitions and of the SiO \jcc\ one in three symbiotic stellar systems
(SSs), R Aqr, CH Cyg, and HM Sge. \doce\ \jdu\ and \juc\ emissions were
detected in R Aqr and CH Cyg. An accurate line profile of \doce\ \jdu\
in R Aqr was obtained.

The observed lines are very weak. If we compare them with those usually
observed in standard evolved stars with similar properties \citep[see
e.g.][]{bujetal92}, the CO lines in SSs are $\sim$ 100 times weaker,
even if we also consider objects, like young planetary nebulae, in
which molecule photodissociation seems to be already efficient.

The weak intensities of the observed lines suggests that CO emission
comes from a very compact region, much smaller than the region
responsible for low-$J$ CO emission in circumstellar envelopes (CSEs)
around standard AGB stars (typical radius larger than 10$^{16}$
cm). This result is compatible with observations (mostly nondetections)
in SSs of other molecular lines, like the SiO, H$_2$O and OH masers,
see Sect.\ 1.
We have shown, from very general considerations, Sects.\ 3, 4, that the
detected CO intensities are coincident with those expected for emission
coming from a circumstellar region comparable in radius to (or slightly
smaller than) the distance between the stellar components of the
system.  We assumed that these molecule-rich layers in SSs have
physical conditions similar to those of the inner layers around
standard AGB stars. Particularly satisfactory is the fitting of the
lines observed in the most intense source, R Aqr, in which the stars
are separated by typically 2.3 $\times$ 10$^{14}$ cm, see Sect.\ 1. 
 
We have also performed model calculations of the CO emission of these
shells, assuming emitting region radii of that order and the physical
conditions and dynamics expected in gas placed at similar distances
around standard AGB stars (Sects.\ 3, 4). Our calculations reproduce
the observed intensities, within the observational uncertainties. We
can also reproduce the \doce\ \jdu\ line profile observed in R Aqr, see
Sect.\ 4.1 and Fig.\ 4. Our model explains in particular the asymmetry
observed in the profile, as a result of selfabsorption in an optically
thick line and in the presence of significant gas acceleration and
systematic temperature decrease with radius. All these properties are
expected to be present in the very inner layers from which we are
assuming that the CO emission comes.

{The data obtained in CH Cyg are worse and their analysis is more
uncertain.}  To explain the observed intensities, we deduce a typical
radius of about 10$^{14}$ cm, again slightly smaller than the distance
between the stars (Sects.\ 1, 4.2). In this region, the physical
conditions seem also comparable to those typical of AGB stars, although
the expansion velocity, $\sim$ 25 \kms, is slightly larger than those
usual in CSEs around standard red giants.

We so conclude that the nebulae around the SSs R Aqr and CH Cyg include
a very compact molecule-rich region, with radii $\sim$ 10$^{14}$ --
2\,$\times$\,10$^{14}$ cm, from which the molecular emission detected
in these objects comes. This region is in extent comparable to or
slightly smaller than the region within both components of the binary
system. We deduce in these regions outwards expansion velocities of
about 5 -- 25 \kms, with a significant acceleration of the gas,
temperatures decreasing with radius between about 1000 and 500 K, and
densities $\sim$ 10$^9$ -- 3\,$\times$\,10$^8$ cm$^{-3}$. The general
physical conditions and dynamics in this inner shell would be then
similar to those typical of standard AGB stars.  This result is
supported by the 'normal' observational properties of SiO maser
emission from R Aqr (Sects.\ 3.4, 4), which show that the CSE around
this star is relatively normal (i.e.\ that its main properties are
comparable to those of standard AGB stars) up to distances $\sim$ 5
10$^{13}$ cm, at least.

{We find relatively high values of the gas mass and mass-loss
rate. In R Aqr, we estimate that the total mass of the CO emitting
region is $\sim$ 2 -- 3 $\times$ 10$^{-5}$ \ms, and that the value of
\mloss\ ranges between 5 10$^{-6}$ and 10$^{-5}$ \my, in agreement with
results obtained from dust emission. These relatively high values
therefore confirm the general trend of SSs to present significantly
higher mass-loss rates than standard, isolated evolved stars
\citep{miko99, miko02}.  We must take into account, however, that the
comparison between mass-loss rates in the very extended and isotropic
winds from standard AGB stars and the values of \mloss\ deduced for our
sources is difficult. Here, we are really detecting gas which is
confined within the orbit of the companion, and its future dynamics is
difficult to know. It is probable that most of this material will never
leave the system (in this case we would not be dealing with a true mass
ejection) or will leave it by means of outbursts or bipolar jets. It is
also possible that the process of mass ejection in these stars is
significantly affected by the companion and cannot be easily compared
with that in isolated AGB stars. Hydrodynamical simulations
\citep{gav02,pods07} show indeed that, in systems with orbital
parameters similar to those of CH Cyg and R Aqr, the radius of the dust
shell is comparable with the Roche lobe radius and a wind Roche-lobe
overflow will occur.

In any case, we think we can safely conclude that the mass of the gas
in the close surroundings of R Aqr and CH~Cyg is relatively high, and
that its presence is probably related to the gravitational effects of
the nearby companion, even if these processes are not well understood. }

We note that the few symbiotic systems showing molecular emission (as R
Aqr and CH Cyg) are objects showing particularly large distances
between the stars \citep{schwarz95}. It is then probable that in other
SSs the molecule-rich region is an extremely small shell tightly
surrounding the Mira-type component, which would explain its very weak
molecular emission.

From the properties we derive for the CO-rich region and the
nondetection of SiO thermal lines (Sect.\ 4.3), whose emitting region
must be significantly smaller than for CO, we suggest that the small
extent of the molecule-rich gas in these sources is mainly due to
molecule photodissociation by the hot dwarf star, which must be
significantly more efficient for low-abundance molecules. Ionized gas
has been found to be very extended in SSs, particularly in the well
studied R Aqr \citep[see e.g.][]{corradi03, hollis99a, hollis99b},
where bipolar and disk-like ionized-gas structures are known to extend
over more than one arcminute (several 10$^{17}$ cm), apparently without
any molecule-rich counterpart.  It is obvious that the relatively
standard inner shells around the red giant show very different
properties than the spectacular outer parts of the nebula. Therefore,
disruption of the CSE due to the strong stellar interaction in SSs must
also play an important role in the confinement of gas (in particular,
molecule-rich gas) in a very compact and dense shell. \cite{seaquist95}
proposed that the powerful wind from the companion may sweep much of
the material ejected by the Mira star. This effect would explain the
strong discontinuity found between the general properties of the
regions inside and outside the orbit of the companion and, following
\cite{seaquist95}, could increase the efficiency of photodissociation
in the rest of the nebula, helping to explain the lack of molecular
emission in SSs.

\begin{acknowledgements}
This work has been partially supported by the Polish Research Grant
No. N203 395534. We acknowledge with thanks the variable star
observations from the AAVSO International Database contributed by
observers worldwide and used in this research. This research has also
made use of the SIMBAD database, operated at CDS, Strasbourg, France
\end{acknowledgements}

\bibliographystyle{aa}

\end{document}